\documentstyle[multicol,pra,aps]{revtex}

\newcommand{\beq}{\begin{equation}}
\newcommand{\eeq}{\end{equation}}
\newcommand{\beqy}{\begin{eqnarray}}
\newcommand{\eeqy}{\end{eqnarray}}

\newtheorem{Lemma}{Lemma}
\newtheorem{Theorem}{Theorem}
\newtheorem{Corollary}{Corollary}
\newtheorem{Example}{Example}
\newenvironment{Proof}{{\it Proof: \,}}{$\Box$}
\newenvironment{Proofw}{{\it Proof \,}}{$\Box$}

\title{Fragility of a class of highly entangled states of 
many quantum-bits}

\author{D.~Janzing\footnote{Electronic address: janzing@ira.uka.de}
 and Th.~Beth\footnote{Electronic address: EISS\_office@ira.uka.de}}
\address{
	Institut f{\"u}r Algorithmen und Kognitive Systeme,
        Universit{\"a}t Karlsruhe, Am Fasanengarten 5, 
	D--76\,128 Karlsruhe, Germany
}

\def\C{{\rm\kern.24em
    \vrule width.02em height1.4ex depth-.05ex
    \kern-.26em C}}
\def\R{{\rm I\kern-.25em R}}
\def\F{{\rm I\kern-.25em F}}
\def\Z{{\rm\kern.26em
    \vrule width.02em height0.5ex depth 0ex}
    \kern.04em
    \vrule width.02em height1.47ex depth-1ex
    \kern-.34em Z}

\begin{document}
%\twocolumn
%\narrowtext
\maketitle
\begin{abstract}

We consider  a Quantum Computer with $n$  quantum-bits
(`qubits'), where each qubit 
is coupled independently to an environment affecting the state in a dephasing or
depolarizing way. 
For mixed states we suggest a  quantification for  the property of showing
{\it quantum} uncertainty on the macroscopic level.
We illustrate in which sense a large parameter can be seen as  an indicator for 
large
entanglement and give hypersurfaces enclosing the set
of separable states.
Using methods of the classical 
 theory of maximum likelihood
estimation we prove  that this
 parameter is
 decreasing with $1/\sqrt{n}$ 
for all those states which have been exposed to the environment.

Furthermore we consider a Quantum Computer with perfect 1-qubit
 gates and 2-qubit gates with
depolarizing error and show that any state which can be obtained from a
 separable initial state  lies inbetween a family of pairs
 of certain hypersurfaces parallel to those enclosing the separable 
ones.
\end{abstract}

\begin{multicols}{2}

%==========================================================================
\section{Introduction}
%==========================================================================
%\onecolumn
%\narrowtext

The sensitivity of quantum systems to interactions with
the environment is one of the challenging problems
for the realization of Quantum Computers \cite{SZL,Un,Di}.
But apart from this motivation the effect of the
 environment to quantum states
has been subject of pure research  for many decades 
(see \cite{GJKKSZ} and references therein).
After all the decoherence caused by the environment is commonly
 accepted to be the explanation for classical behavior of physical
systems in every-day life \cite{Zu1,Zu3,Pr}, i.e., on the {\it macroscopic}
level of physics \cite{GJKKSZ,Zu2}.
Although there is no precise definition  of the  word `macroscopic',
most of those explanations contain (explicitly or implicitly)
the statement that the destruction of coherence takes place
on a very small time scale particularly for 
superpositions of `macroscopically distinct' states, i.e.,
States showing {\it quantum} uncertainty on the macroscopic level. Despite
the fact, that this statement cannot be maintained 
without taking into account the way of coupling to the environment
(see \cite{BBH}), it has served as an intuitive motivation 
for our investigation of the sensitivity of many particle quantum states
with respect to a coupling to {\it independent} environments.
For this 
 we introduce a function $e: \rho\mapsto e_\rho$ from the set of states
to the positive numbers
 quantifying the property
of showing quantum uncertainty on the macroscopic level and
  prove quantitative statements about the sensitivity
of those states $\rho$ having  large values $e_\rho$.

We show that large values for $e_\rho$
require large-scale-entanglement in the sense that there cannot be
{\it small} clusters of entangled qubits without entanglement between
qubits in different clusters.
Therefore our investigations should be considered
in the context of the fragility of entanglement, which is
an important subject since it is decisive for
the computational power of decohered Quantum
 Computers \cite{AB}.

The connection between the sensitivity of a state
with respect to disturbances of the environment and the property of
being a superposition of macroscopically distinct states
can be illustrated by the following straightforward example:
Take an $n$-qubit Quantum Computer, that is a quantum system with
 the Hilbert space 
\[
\underbrace{\C^2\otimes\dots \otimes \C^2}_{n}.
\]
 Denote the
canonical basis states by the binary words of length $n$.
Furthermore we use the following terminology:
The Hamming
{\it weight} of a binary word is the number of characters `1'.
The Hamming {\it distance} of two words is the 
Hamming weight of their difference.
Now we  take a superposition of two arbitrary basis states
\[
|\psi\rangle:= \frac{1}{\sqrt{2}}(|a\rangle +|b\rangle).
\]
If $a$ and $b$ differ at many positions, i.e., have a large Hamming distance,
 we call $|a\rangle$ and $|b\rangle$ 
{\it macroscopically} distinct states.
Now we perform a measurement of $|\psi\rangle$ 
in the canonical basis $|0\rangle, |1\rangle$ 
of $\C^2$ on
one  randomly chosen qubit $i$.
Obviously this state collapses to 
$|a\rangle$ or $|b\rangle$ if and only if the words $a$ and $b$ differ
 at the position $i$.
Hence the superposition state is more {\it fragile} if $a$ and $b$  have a
 large Hamming distance.
At the first sight it seems to be straightforward to characterize
the fragility of a state (with respect to dephasing) by
the probability that it is changed by a measurement of a
 randomly chosen qubit.
But this  probability is 1 in the generic case:
Even the unentangled state 
\[
|\pi\rangle:=(\frac{1}{\sqrt{2}}(|0\rangle +|1\rangle))^{\otimes n}
\]
is changed by a measurement of any qubit\footnote{
In \cite{Un} this state is taken as an example
for the difficulty of maintaining coherence in
large Quantum Computers.}.
Nevertheless  we want to consider this state
as much less fragile than the `cat-state'
\[
|\gamma\rangle:=\frac{1}{\sqrt{2}}(|0\dots 0\rangle +|1\dots 1 \rangle)
\]
 since the `error' caused by the  
single-qubit-measurement can be corrected by a single-qubit-operation
 in the case of the state $|\pi\rangle$ whereas 
the local disturbance of the cat state $|\gamma\rangle$ requires 
a much more  complicated procedure restoring the entanglement.
Therefore we   consider fragility 
as a property of a {\it class of states} rather than of a single state.
We prove a class of highly entangled states to be fragile in the sense that
every mixed state obtained by small independent disturbances
of each qubit lies outside this set.

\section{Two sufficient conditions for entanglement
 in many particle systems}\label{entanglement}

As usual
we call a state of the Quantum Computer a {\it product state} if its density 
matrix is an $n$-fold tensor product of the form
\[
\rho_1\otimes \rho_2 \otimes \dots \otimes \rho_n
\] 
where each $\rho_i$ is the density matrix of the qubit $i$.
A {\it separable} state is a convex combination of arbitrary many
 product states.
For any operator $a$ acting on an Hilbert space we
denote its operator norm by $\|a\|$.
By an `1-qubit-operator  at the qubit $i$' (or `acting on the qubit $i$')
 we mean an operator acting on the $n$-fold tensor product
of $\C^2$ which is of the form
\[
\underbrace{1\otimes 1\otimes\dots \otimes}_{i-1}  a
 \underbrace{\otimes \dots 1\otimes 1}_{n-i}
\]
for any $a\neq 1$ acting on $\C^2$.

Furthermore we introduce the following type of observables:
Let $(a_i)_{1\leq i\leq n}$ with $\|a_i\|\leq 1$  be a family
 of  selfadjoint operators where
each $a_i$ is  acting on the
 qubit $i$. Then we define the {\it averaging observable}
\[
\overline{a}:=\frac{1}{n}\sum_i a_i.
\]
In the case that the difference of the lowest and the greatest eigenvalue
is the same for every $a_i$ we call $\overline{a}$
an {\it equally weighted average} of $1$-qubit observables.
Despite the fact, that 
there is no precise distinction between
the macroscopic and microscopic level in an $n$-qubit system, 
we have good reasons for  considering the equally weighted averages as
the `most macroscopic ones'.
An easy example might illustrate this point of view:
 If the qubits are represented
as spin-1/2-particles, the mean-magnetization of the system in 
 z-direction is given by the averaging observable obtained by
  setting
\[
a_i:=\sigma^{(i)}_z,
\]
where $\sigma^{(i)}_z$ is the copy of the Pauli matrix $\sigma_z$ acting 
on the  qubit $i$. If we would  define
\[
a_i:=\lambda_i \sigma^{(i)}_z
\]
with arbitrary $\lambda_i$ we get a {\it less} macroscopic
observable in general since $\overline{a}$ is dominated by 
the spins of those $i$
with large $| \lambda_i |$.    

In a product state  there is no correlation
between the values of two 1-qubit observables at different qubits. Hence
we conclude the following from classical probability theory:

For any observable $a$ and any density matrix $\nu$
let $s_{a,\nu}$ be the standard deviation of 
$a$ in the state $\nu$, i.e.,
\[
s_{a,\nu}:=
\sqrt{tr (\nu a^2)-(tr (\nu a))^2}.
\]
Let $\rho$ be a product state.
   Then for the standard deviation 
of any averaging  observable $\overline{a}$ in the state $\rho$
the inequality
\beq\label{sigma}
s_{\overline{a},\rho}\leq \frac{1}{\sqrt{n}}
\eeq
holds since the variance of a sum of independent random variables
is the sum of their variances and the variance
of any $a_i$ cannot exceed 1 due to its operator norm. 
We conclude, that every separable state has a decomposition 
into states fulfilling inequality (\ref{sigma}) for every family 
$(a_i)$.
If we want to use this result for showing that a given state is not separable
one would have to check every possible decomposition into pure states. 
Hence
one might doubt its practical importance.
However, we can derive another sufficient condition for entanglement
which does not require to check every decomposition:
The question, as to which extent  a state can be decomposed into those pure 
states
with small standard deviations with respect to any given observable $a$,
is closely related to the question as to which extent its density matrix
(written in any basis diagonalizing $a$) is dominated
by `strongly' off-diagonal positions, i.e., those positions
where row and column correspond to rather different eigenvalues of $a$.
The convex function
\[
\rho \mapsto \sup_{\|b\|\leq 1} |tr (\rho [a,b])|
\]
can be considered as a measure for the `dominance of the
strongly off-diagonal' terms since it vanishes for every $\rho$ commuting
 with $a$.  

To be precise, we have the following lemma:

\begin{Lemma}\label{sigma2}

Let $A$ be an arbitrary (finite dimensional) matrix algebra.
Let $b,c \in A$ with $b$ selfadjoint and $\|c\|\leq 1$.
Let $\nu\in A$ be an arbitrary density matrix.
Then we have the following  inequality:
\begin{equation}\label{commutator}
|tr (\nu [b,c])|\leq 2 s_{b,\nu}
\end{equation}
\end{Lemma}

\begin{Proof}
Since the standard deviation is a concave function on the set
of probability measures on $\R$,
we have $s_{b,\nu}\geq \sum_j\lambda_j s_{b,\nu_j}$ if $\nu$ is
the convex sum $\nu:=\sum_j \lambda_j \nu_j$.
Therefore we can assume $\nu$ to be
a density matrix of a pure state,
  i.e. $\nu=|\psi\rangle\langle
  \psi |$.
Let us choose an eigenvector basis of $b$ and expand $c$ and $|\psi\rangle$
 with respect to this basis.
 Let $\lambda_1,..,\lambda_k$
be the set of eigenvalues of $b$ with their
 corresponding multiplicities and $\psi_j$ be the coordinates
of $|\psi \rangle $.
We have:
\beqy\label{kappagl}
&&|\langle \psi |[b,c] \psi \rangle |= |\sum_{ij} \psi_i
 (\lambda_i-\lambda_j)
\overline{c}_{ij}\overline{\psi}_j|\nonumber\\
& \leq& 
|\sum_{ij} \psi_i (\lambda_i -\mu)  \overline{c}_{ij}
 \overline{\psi}_j|+
|\sum_{ij}\psi_i  \overline{c}_{ij} (\lambda_j-\mu)
 \overline{\psi}_j   |,
\eeqy
for any $\mu\in \R$.

Defining the vector $\kappa$ with 
$\kappa_i:=(\lambda_i-\mu)\psi_i$.
 the inequality (\ref{kappagl}) reads  
\[
|\langle \psi |[b,c]\psi\rangle| \leq 
|\langle \kappa | c \psi\rangle | + |\langle\psi | c\kappa\rangle|,
\]
which gives us  an upper bound by the Cauchy Schwartz inequality
\[
|\langle \psi |[b,c]\psi\rangle| \leq 2\|\kappa\| \, \|c\psi\|\leq 2\|\kappa\|,
\]

where the last estimation holds due to the operator norm of $c$.
With the definition  $\mu:=\langle \psi | b \psi\rangle$ 
the vector norm
$\|\kappa\|$ is the standard
 deviation of the observable $b$.
\end{Proof}

By the triangle inequality we conclude:

\begin{Corollary}\label{gewichteteStandard}
Let $\rho$ be a density matrix of a finite dimensional quantum system.
Let $b$ be an arbitrary selfadjoint operator and $c$ be an arbitrary
 operator
with $\|c\|\leq 1$.
Assume $\rho$ to have a decomposition into pure states of the form
\beq\label{dec}
\rho=\sum_{j} \lambda_j \rho_j,
\eeq
where $\lambda_j >0$ and $\sum_j\lambda_j=1$. The states  $\rho_j$
 are arbitrary 
pure states.
Then we call   $\sigma:=\sum_j s_{b,\rho_j}$ `the mean standard deviation
 of the observable $b$ in the state
$\rho$  with respect to the decomposition  
(\ref{dec})' and
 have the following inequality:
\[ 
\frac{1}{2}|tr (\rho [b,c])|\leq \sigma
\]
\end{Corollary}

Note that for any pair of selfadjoint operators $b,c$ and any state $\nu$ the 
expectation value $i\, tr( \nu [b,c])$ is real.
For any such pair $b,c$ we define the hypersurface
\[
H_{b,c,r}:=\{\rho \in {\cal S} \, | \, i\, tr(\rho [b,c])=r
\}
\]
where  ${\cal S}$ is  the set of density matrices of the $n$-qubit 
system. 
We conclude from Corollary \ref{gewichteteStandard}:

\begin{Corollary}(`Hypersurface-Criterion')
Let  $\overline{a}$ be an averaging observable
 as in the beginning of this section 
 and $c$ be a selfadjoint operator with $\|c\|\leq 1$.
   Then every separable state $\rho$
lies inbetween the hypersurfaces 
$
H_{\overline{a},c,\pm \frac{2}{\sqrt{n}}},
$
i.e., 
\beq\label{hypers}
-\frac{2}{\sqrt{n}}\leq i\, tr(\rho [\overline{a},c]) \leq \frac{2}{\sqrt{n}}   
\eeq
\end{Corollary}

This  gives us a sufficient condition
for entanglement which is easy to handle since it can be verified by
finding just one pair $\overline{a}, c$ such that inequality \ref{hypers}
is violated.

Despite the fact, that there is no commonly accepted quantification of 
entanglement (see \cite{VPRK,SM1,SM2}), we will
consider  a large value (compared to 
$1/\sqrt{n}$) of the term
 $tr(\rho [\overline{a},b] )$ 
for any  such family $(a_i)$ and any such  $b$
as a {\it sufficient} condition for $\rho$ to be  `highly entangled'.
The term {\it highly} entangled might be interpreted 
in two different ways: Firstly a large value
 shows that the state
has a great distance from
the set of separable states in the trace norm.
Secondly it shows that there is entanglement between {\it many}
 qubits:

\begin{Lemma}
Take a partition of the qubits $\{1,\dots,n\}$ into
subsets (`clusters') of size   
 $l_1,\dots,l_k$ with the property that the state 
$\rho$ has no entanglement
between qubits of different clusters.
Let $\overline{a}$ be an averaging observable  and $b$ be
a selfadjoint operator with $\|b\|\leq 1$. Then
 the following inequality holds:
\beq\label{cluster}
|tr (\rho [\overline{a},b])|\leq \frac{2}{n}\sqrt{\sum_{i\leq k} l^2_i}
\eeq
\end{Lemma}

\begin{Proof}
Denote the clusters by $S_1,\dots, S_k \subset \{1,\dots, n\}$.
It is sufficient to take a state which is
 factoring with respect to this partition into clusters since 
the set of states fulfilling inequality (\ref{cluster}) is convex.
For such a `partial product state' $\rho$
the standard deviation of $\overline{a}$ is given
by 
\[
s_{\overline{a},\rho}=\frac{1}{n}\sqrt{\sum_{i\leq k} s^2_i}
\]
 where $s_i$ denotes the standard 
deviation of the observable $\sum_{j\in S_i} a_j$ which is less or equal
 to $l_i$. Lemma \ref{sigma2} completes the proof.
\end{Proof}

In the following, we will restrict ourselves to  the 
equally weighted averaging observables.
 Up to a constant factor and a constant
summand they can be obtained by taking every $a_i$ as a projection. 

Therefore,
 we shall consider
the parameter
\beq\label{paradef}
e_\rho:=\sup_{\overline{Q},b} |tr (\rho [\overline{Q},b])|
\eeq
where every $\overline{Q}$ is
the average $\overline{Q}$ over the 1-qubit
projections $Q_k$ and $\|b\|\leq 1$, 
as a reasonable quantification for the property of showing
{\it quantum} uncertainty on the macroscopic level.

\begin{Example}\label{ex}
In order to give more intuition about the states with large $e_\rho$,
we assume  $\rho$ to be the density matrix of
a coherent superposition of two distinct basis states $|f\rangle$
and $|g\rangle$, i.e., $\rho=\frac{1}{2}(|f\rangle +|g\rangle)(\langle 
f|+\langle g|)$.  Let  $f$ and $g$ be binary words with Hamming
weights
 $wgt(f)$ and $wgt(g)$.
Let $P_k$ be the projection onto the state $|1\rangle$ 
for the $k$-th qubit.
With the definition
\[
b:=i|f\rangle \langle g|-i|g\rangle\langle f|.
\]
 we get
\[
 tr ( \rho [\overline{P},b])=\frac{i}{n}\,(wgt(f)-wgt(g)),
\]
and hence we have $e_\rho\geq \frac{1}{n} | wgt(f)-wgt(g)|$.
\end{Example}
We prove the following more general statement:

\begin{Lemma}
Let $f_1,\dots f_j, g_1,\dots, g_j$ be a set of $2j$ distinct binary words.
Let $\rho$ be the density matrix given by
\[
\rho:=\sum_k \lambda_k |\psi_k\rangle\langle \psi_k| 
\]
where $\lambda_k$ is the probability of the pure superposition 
state
\[
|\psi_k\rangle:=\frac{1}{\sqrt{2}}(|f_k\rangle + |g_k\rangle).
\]
Define $b$ by
\[
b:= \sum_k( i|f_k\rangle\langle g_k| -i |g_k\rangle\langle f_k|).
\]
Then we have the following equation:
\[
tr (\rho [\overline{P},b]) =i\frac{1}{n} \sum_k \lambda_k (wgt(f_k)-wgt(g_k)).
\] 
\end{Lemma}

\begin{Proof}
Using
\[
\overline{P}|f\rangle =\frac{1}{n}wgt(f)|f\rangle
\]
for every binary word $f$, the statement follows by easy calculations.
\end{Proof}

Note that the operator $b$ in the definition  above fulfills the
 requirement  $\|b\|=1$ since the operators 
\[
i|f_k\rangle\langle g_k|
-i|g_k\rangle \langle f_k|
\]
 have operator norm 1 and 
act on mutually orthogonal 
subspaces. 

The states $|\psi_k\rangle$ 
are superpositions of macroscopic distinct states
if the difference $wgt(f_k)-wgt(g_k)$ has the order of $n$ rather
 than the order of $1$.
In this case we say the state  shows quantum uncertainty
on the macroscopic level.
The sum $\sum_k \lambda_k (wgt(f_k)-wgt(g_k))$ measures
to what extend the mixture $\rho$  consists of pure states
with a large uncertainty of the observable $\overline{P}$.
Note that the mixture of two states with large parameter $e$ can have
small $e$ due to the fact that the mixture of two highly entangled
states can be separable. Therefore
the assumption that the  $2j$ binary words
$f_1,\dots, f_j,g_1,\dots, g_j$ are mutually distinct is essential 
and turns up not to be just a technical requirement for the proof:
Take the mixture given by 
\[
\rho:=\frac{1}{2}(|\psi_1\rangle\langle \psi_1|+|\psi_2\rangle\langle \psi_2|)
\] 
with
\[
|\psi_{1/2}\rangle :=\frac{1}{\sqrt{2}}(|0\dots 0\rangle \pm |1\dots 1\rangle).
\]
Easy calculation 
in the canonical basis  shows
that   $tr( \rho [\overline{P},b])$ 
vanishes for every operator $b$ since $\rho$ and $\overline{P}$ are diagonal
in this basis.
Actually, we can get this result using Lemma \ref{sigma2}  as well:
The state $\rho$ has another decomposition into the pure states
$\rho_0:=|0\dots 0\rangle\langle 0\dots 0|$ and 
$\rho_1:=|1\dots 1\rangle \langle 1\dots 1|$.
Both states do not show any uncertainty with respect to the observable
$\overline{P}$, i.e., the standard deviations  $s_{\overline{P},\rho_0}$ and 
$s_{\overline{P},\rho_1}$ vanish. 

We can make general statements about the range of the convex function
 $\rho \mapsto e_\rho$ on the set of density matrices:

\begin{Lemma}
For every $n$ the range of the function $ \rho \mapsto e_\rho$
is the interval $[0,1]$.
\end{Lemma} 

\begin{Proof}  
If $\rho$ is the maximally mixed state, i.e., $\rho$ is
the identity matrix up to a constant factor, we have $e_\rho=0$ since
$tr(\rho [a,b])=tr([a,b])=0$ for every pair of operators $a,b$.
In general we have $e_\rho\leq 1$ due to Lemma \ref{sigma2} since
 the standard deviation of $\overline{P}$
cannot exceed $1/2$ due to the fact that its spectrum  is contained
in  $[0,1]$.
For the cat state as defined in the introduction we can conclude
$e_\rho\geq 1$ by setting $f:=0\dots0$ and $g:=1\dots 1$ in 
Example \ref{ex}.
Hence for  the cat state we have $e_\rho=1$.
We can obtain
any  value between $0$ and $1$ by a mixture 
of the cat state 
and the maximally mixed state with the corresponding weight.
\end{Proof}

\section{The error models}\label{error}

In many cases it is well-justified from a physical point of view
to assume errors acting independently on every qubit \cite{AB}.
One kind of these 1-qubit-error which seems reasonable
is a random dephasing with respect to the canonical basis.
Describing this on the set of density matrices this error
affects the state like a measurement instrument.
We describe the effect of the dephazing environment
by a map 
 $G$ on\footnote{$G$ is 
 a completely positive trace preserving  map (see \cite{AKN}), since every
 manipulation of a quantum state
can be described by a map of this type.}
 the set of $n$-qubit density matrices as follows:

Let $P_i$ be as in Example \ref{ex} and
 $M_i$ the instrument performing a measurement 
of the qubit $i$
in the canonical basis of $\C^2$, i.e.,
\[
M_i(\rho)=P_i \rho P_i + (1-P_i)\rho (1-P_i)
\]
for every density matrix $\rho$.
 Then our first error model
will be an instrument $G$  which acts on each qubit independently 
as $wM_i+ (1-w)id$,  i.e.  $G$ acts as:
\[
G:= \prod_i (wM_i+(1-w)id)
\]
where $id$ denotes the identity map
and $w$ is the error probability.

Our second error model is given by depolarizing
 channels acting on each
qubit independently:
Let $I$ be this map on the set of 1-qubit density matrices which maps
every state to the maximally mixed one.
Let $I_i$ be the canonical extension of this map from the state space
of the qubit $i$ to the $n$-qubit-system, i.e.,
\[
I_i:= id \otimes \dots \otimes id \otimes I\otimes \dots \otimes id.
\]
Define the instrument $D$ by:
\[
D:=\prod_i (wI_i+(1-w)id).
\]
In the following chapter we shall study the images of the maps $D$ and $G$
according to the error probability $w$ and the size $n$.

\section{Quantitative statements about fragility}

In order to investigate the way in which the instrument $G$
affects a state we 
introduce a family of instruments
$(G_l)_{l\leq n}$ which is defined as follows:
Let ${\cal L}_l$ be the set of l-element subset of $\{1,\dots,n\}$.
Set
\[
G_l:=\frac{1}{\left(\begin{array}{c} n\\l\end{array}\right)}
\sum_{L\in {\cal L}_l} \prod_{i\in L}M_i.
\] 
Then easy calculation shows that $G$ can be written as the convex
 combination of all the $G_l$ with binomial
 coefficients: 
\beq\label{Gdef}
G=\sum_{l=0 }^n B_{nw}(l)\, G_l.
\eeq
where we use the abbreviation
\[
B_{nw}(l):= \left( \begin{array}{c} n \\ l \end{array}\right)  
w^l (1-w)^{n-l}.
\]
The instrument $G_l$ is a
 random machine performing a measurement
on every qubit in a randomly selected l-element subset of qubits.
It can map a pure state to a mixed one for two 
reasons: Firstly we do not know, which $l$ qubits are measured, i.e.,
which $L$ was selected,  
and secondly we do not know the measured result. 
If we knew both, we would  get a certain pure state obtained from 
the original
one by a partial collapse of the wavefunction.
 In the following we show, that
most likely this collapse leads to a  state in which
the observable $\overline{P}=\frac{1}{n}\sum P_i$ has small standard 
deviation provided that $l>>1$. Intuitively this is not astonishing since 
the measurement of $l$ qubits allows a prediction of the
values of the
average observable $\overline{P}$ with a high
 `confidence level'.  
This analogy to the theory of  {\it maximum likelihood
 estimation}
motivates the idea of the proof in
 the following quantitative analysis:

\begin{Theorem}\label{GTheorem}  
Let $\rho$ be an arbitrary density matrix of an $n$ qubit system. 
Let $G_l$ and $\overline{P}$ as described above.
For $b$ an arbitrary  operator on $(\C^2)^{\otimes n}$
 with $\|b\|\leq 1$ we have 
\[
|tr (G_l(\rho) [\overline{P},b])| \leq \frac{1}{\sqrt{l}} 
\sqrt{\frac{n-l}{n-1}}.
\]
\end{Theorem}

Before we prove the theorem we draw some conclusions.
In order to get statements about $G(\rho)$ instead of $G_l(\rho)$
we can use the decomposition (\ref{Gdef}). For this we have to consider the case
$l=0$ separately. The instrument $G_0$ is the identity map and 
occurs in the sum (\ref{Gdef}) with the weight $(1-w)^n$.
 The  standard deviation of $\overline{P}$ can never exceed $1/2$.
Therefore,
 $|tr(G_0(\rho)[\overline{P},b])|=|tr(\rho [\overline{P},b])|\leq 1$
and we conclude:

\begin{Corollary}\label{Gungl}
Let $G$ be
as in equation (\ref{Gdef}).
 Then we have for any arbitrary density matrix $\rho$
\beqy\label{Gungleq}
|tr (G(\rho)[\overline{P},b])|&\leq&  \sum_{l=1}^n 
\frac{1}{\sqrt{l}}\sqrt{\frac{n-l}{n-1}} B_{nw}(l)   \\
&+&(1-w)^n\nonumber\\
&=:&r_{wn}\nonumber .
\eeqy
\end{Corollary}

\begin{Proofw} (of the Theorem):
Since the set of density matrices fulfilling the inequality is convex,
we may restrict the proof to the case of $\rho$ being 
 a pure state, i.e. $\rho=|\psi\rangle
\langle \psi |$ with $|\psi \rangle \in (\C^2)^{\otimes n}$.
Let us assume a measurement which has been performed on the qubits in $L$,
 where
$L$ is an arbitrary $l$-element subset of $\{1,\dots,n\}$.
Let $R_L$ be the map
 $R_L:\{0,1\}^n\rightarrow \{0,1\}^l$ restricting
a binary word to the set $L$. 
Let $P_{L,g}$ be the projector onto the 
linear span of those basis vectors given by the
 binary words in  $R_L^{-1}(g)$ for any $g\in \{0,1\}^l$.

Then  $|\psi\rangle \langle \psi|$ is transduced to the mixed state
\[
\sum_{g\in \{0,1\}^l} P_{L,g}|\psi\rangle \langle \psi| P_{L,g}.
\]

On the set  ${\cal L}_l$  we
introduce the measure $r_l$ as the equally distributed
 probability measure,
i.e. 
\[
 \forall 
L\in {\cal L}_l: \, r_l(L)=\frac{1}{\left(\begin{array}{c} n \\l 
\end{array}\right)}. 
\]

Then $G_l$ transduces
$|\psi\rangle\langle \psi |$ to
the state
\beq\label{Geinfach}
G_l(|\psi\rangle \langle \psi|)=
\sum_{L\in {\cal L}_l} r_l(L)
 \sum_{g\in \{0,1\}^{l}}P_{L,g}
|\psi\rangle \langle  \psi| P_{L,g}.
\eeq

With the definition $|\psi_{L,g}\rangle:=|P_{L,g}\psi\rangle/
\|P_{L,g}\psi\|$
and
\beq\label{pDef}
p(L,g):=r_l(L)\|P_{L,g}\psi\|^2.
\eeq
we obtain:
\[
G_l(|\psi\rangle\langle \psi |)
=\sum_{L\in {\cal L}_l, g\in \{0,1\}^l}p(L,g) |\psi_{L,g}\rangle\langle
 \psi_{L,g}|.
\]

Note that $p(L,g)$ is the probability for the event 
`a measurement has been performed
on the qubits in $L$ and the result $(g_1,\dots,g_l)$ (in an ascending order)
has been obtained.' 
We denote this event by $(L,g)$.

Let $s_{L,g}$ be the standard deviation of $\overline{P}$ in the state
$|\psi_{L,g}\rangle \langle \psi_{L,g}|$. 

In order to prove the theorem it is sufficient (see Corollary
 \ref{gewichteteStandard}) to show

\beq\label{sum}
\sum_{L,g} p(L,g) s_{L,g} \leq  \frac{1}{2\sqrt{l}} \sqrt{\frac{n-l}{n-1}}.
\eeq

For doing so we introduce the probability
 space 
\[
{\cal L}_l\times \{0,1\}^n
\]
endowed with the product measure $r_l\otimes q_\psi$, where
$q_\psi$ assigns to every binary word the probability given 
 by the square of the probability amplitudes of $\psi$.
All the random variables which will be introduced below
are defined on this product space.
In the formal framework of this space the
formal correct notation of $(L,g)$ is
  $\{L\}\times R_L^{-1}(g)$. We will keep the less
formal notation $(L,g)$ for reasons of convenience.
Now we introduce the random variable 

\[
U: {\cal L}_l\times \{0,1\}^n\rightarrow \R
\]
by
\[
U(L,b):=\frac{1}{n}wgt (b).
\]

Then we can write the standard deviation in inequality (\ref{sum}) as
\[
s_{L,g}= \sqrt{ E((U-E(U|L,g))^2 |L,g) },
\]
where $E( . | L,g)$ denotes the expectation value of a
 random variable
with the {\it conditional} probability measure given the
 event $(L,g)$.
Furthermore we define the random variable $S$ by 
\[
S(L,b):=\frac{1}{l} wgt(R_L(b)).
\]
Due to the fact, that $S$ has a constant value on every subset $(L,g)$,
we can give the following upper bound:

\[
s_{L,g}\leq \sqrt{ E( (U-S)^2 |L,g) },
\]
because for any arbitrary random variable $X$
and $\mu\in \R$ the expectation
value $E((X-\mu)^2)$ is minimized by $\mu=E(X)$.

Since the square root function is concave we get

\beqy
\sum_{L,g} p(L,g)s_{L,g} &\leq& \sqrt{ \sum_{L,g} p(L,g) E (( U-S)^2 |L,g)}\\
&=&\sqrt{\sum_{b\in \{0,1\}^n} p(b) E((U-S)^2 | b)}
\eeqy
The last equation holds since the family of sets 
\[
(L,g)_{L\in {\cal L}_l, g\in\{0,1\}^l}
\]
as well as $b\in \{0,1\}^n$  define different partitions of the probability 
space
${\cal L}_l\times \{0,1\}^n$.
We conclude
\beq\label{hyper}
\sum_{L,g} p(L,g) s_{L,g} \leq \sup_{b\in \{0,1\}^n}\sqrt{E((U-S)^2|b)}.
\eeq
Note that for any fixed $b$ the product 
$Sl$  is a random variable with a hypergeometric distribution, since
it measures the hamming weight of the restriction of $b$ to a randomly
chosen $l$-element subset of $\{1,\dots, n\}$.
Furthermore $U$ is a constant for any given $b$ and $U$ is the expectation
 value of the random variable $S$ with respect to the conditional
probability measure given the event $b$. Therefore the square root in the right
hand term is the standard deviation $\sigma$ of $S$ and is given by (\cite{JK}, 
Sec. 2.3): 
\[
\sigma=\sqrt{\frac{n-l}{l(n-1)} pq}
\]
with  $p:=wgt(b)/n$ and $q:=1-p$.
Since $pq\leq 1/4$ independent of $wgt(b)$  we can estimate
the term (\ref{hyper}) by
\[
\sum_{L,g} p(L,g)s_{L,g}\leq \frac{1}{2\sqrt{l}}\sqrt{\frac{n-l}{n-1}}.  
\]
\end{Proofw}

Since $G$ and $G_l$ describe error models which are not  invariant
with respect to local unitary transformations, 
there is no evident generalization for any other family 
 $(Q_i)$ of projections  instead of $(P_i)$. In contrast, the error 
model
defined by the map $D$ (see end of section \ref{error}) is
 symmetric with respect to the
group $SU_2\otimes SU_2\otimes \dots \otimes SU_2$ of
independent local unitary transformations on every qubit.
Therefore we obtain estimations for the standard deviations
of any observable obtained by averaging over an arbitrary 
family $(Q_i)$ of $1$-qubit-projections and get:

\begin{Theorem}\label{main}
Let $\rho$ be an arbitrary state of an $n$-qubit
Quantum Computer. 
Let $r_{wn}$ as in (\ref{Gungleq}).
Then we have the following inequality:
\[
e_{D(\rho)}\leq r_{wn}.
\] 
\end{Theorem}

\begin{Proof}
Let $\sigma^{(i)}_{x}$ be the
operator representing the Pauli matrix $\sigma_x$ acting on the
 qubit $i$. Let $M_i$ be as in section \ref{error}
and $F_i$ be the instrument performing the bit-flip
$\nu \mapsto \sigma^{(i)}_{x} \nu \sigma^{(i)}_{x} $
on every density matrix $\nu$. 

Then $I_i$ of section \ref{error} is given by:
\[
I_i= (\frac{1}{2}(F_i +id)\circ M_i).
\]
Hence $D$ can be written as the product
\[
D=\prod_{i\leq n} D_i,
\]
 where $D_i$ is  defined by:
\[
D_i=((\frac{1}{2}(F_i+id)\circ M_i w) +(1-w)id).
\]
In analogy to equation (\ref{Gdef}) we can decompose $D$  
into a convex sum of instruments $D_l$ where
$D_l$ is a  machine performing a depolarizing error on a
randomly chosen $l$-element subset $L \subset \{1,\dots,n\}$ of qubits.
Hence we have:
\beqy\label{Dl}
D_l (|\psi\rangle\langle \psi|)&=&\sum_{L\in {\cal L}_l} r_l(L) 
\prod_{i\in L} \frac{1}{2}(F_i+id)\circ M_i (|\psi\rangle \langle \psi|)\\
&=&\sum_{L\in {\cal L}_l} r_l(L) \prod_{i\in L} \frac{1}{2}(F_i+id)\prod_{i\in 
L} M_i (|\psi\rangle\langle \psi|).
\eeqy
Using 
\[
\prod_{i\in L} M_i (|\psi\rangle\langle \psi|)=\sum_{g\in\{0,1\}^l}
\|P_{L,g}\psi\|^2 |\psi_{L,g}\rangle \langle \psi_{L,g}|
\]
and the definition (\ref{pDef}) we obtain
\beqy
&&D_l (|\psi\rangle\langle\psi |)\nonumber\\
&=&\sum_{L\in {\cal L}_l, g\in \{0,1\}^l}
p(L,g) \prod_{i\in L} \frac{1}{2}(F_i+id) (|\psi_{L,g}\rangle\langle
 \psi_{L,g}|)\\
&=& \sum_{L\in {\cal L}_l, g\in \{0,1\}^l}
p(L,g)                                   
 \sum_{T\subset L}\frac{1}{2^l}\prod_{i\in T} F_i(|\psi_{L,g}\rangle\langle 
\psi_{L,g}|) 
\eeqy

This completes the proof:
For any $T$ the standard deviation of $\overline{P}$
in the pure state $\prod_{i\in T} F_i (|\psi_{L,g}
\rangle\langle \psi_{L,g}|)$ is the same as in the
 state $|\psi_{L,g}\rangle \langle
\psi_{L,g}|$, since every $F_i$ is only a permutation of the states $|0\rangle$
and $|1\rangle$ in the qubit $i$.

Therefore we have shown, that $D_l(|\psi\rangle\langle \psi|)$
has a decomposition such that the corresponding mean standard deviation
of $\overline{P}$ is less or equal to 
\[
\frac{1}{2}\sqrt{\frac{n-l}{l(n-1)}}.
\]
Therefore the mean standard deviation of $D(|\psi\rangle\langle\psi|)$
is less or equal to $r_{wn}/2$ (see the  Definition of $r_{wn}$ in 
 (\ref{Gungleq}) and the decomposition of $D$ into a convex sum
of $D_l$) and hence
\[
|tr (D(\rho) [\overline{P},b])|\leq r_{wn} 
\]
by Corollary \ref{gewichteteStandard}.
Due to the symmetry of the error model with respect to local unitary
transformations we can substitute $\overline{P}$ by any other
average $\overline{Q}$ over 1-qubit-projections.
\end{Proof}

The theorem shows, that the `entanglement-parameter' $e_\rho$
of any state $\rho$ is extremely sensitive  to small 
depolarizing perturbations acting on every qubit independently.

At the first sight, the only quintessence of these results seems to be
that they support the well-known fragility of entanglement by
a quantitative analysis without taking into account
the possibility of error correction \cite{St}.
However, we can make easy conclusions for the following
model which is so general as to include every possible error correction 
procedure.
We take a Quantum Computer which allows only an imperfect implementation
of gates. Taking into account that every error correction
has to rely on these gates producing new errors, we see
that there are states which never can be obtained.

The following theorem shows this statement
quantitatively. Therefor
we assume that our Quantum Computer is endowed with the following
set of operations on the set of density matrices:
\begin{itemize}
\item
perfect 1-qubit gates, i.e., maps of the type
$\rho \mapsto u\rho u^*$ where $u$ is an arbitrary unitary operator acting on
1 qubit and $\rho$ is the density matrix of the Quantum Computer
\item imperfect 2-qubit gates $g$ of the following type:
\[
g(\rho):=(1-w)u\rho u^* +w (I_i\circ I_j)(\rho),
\]
where $w$ is the error probability and $u$ is an arbitrary unitary operator
acting on the qubits $i$ and $j$.
\end{itemize}

\begin{Theorem}\label{Haupt}
Let the Quantum Computer (endowed with the basic operations above)
 be initialized in a separable state.
Let $(Q_i)$ be a family of projections where $Q_i$ is acting on the qubit $i$.
Let  $b$ with $\|b\|\leq 1$ be a selfadjoint operator and $r_{wn}$ as in
Corollary \ref{Gungleq}. 

Then it is not possible to prepare a state outside
the slice described by the  pair of hypersurfaces 
\[
H_{{\overline{Q}},b,\pm x}
\]
with 
\[
x:=\frac{1}{n}\sup_{k\leq n}\{ r_{wk} k+\sqrt{n-k}\},
\]
i.e., it is not possible to prepare a state $\rho$ with
$e_\rho> x$.
\end{Theorem}

\begin{Proof}
Let the initial state  be a pure  product state.
Let $k$ be the number of those qubits which are accessed by 2-qubit-gates 
during
the preparation procedure. Without loss of 
generality assume them to be the qubits $1,\dots, k$.
Let $\rho$ be the state obtained by the preparation.
Then we have:
\[
|tr (\rho [\overline{P},b])|\leq \frac{k}{n}
|tr (\rho [\frac{1}{k}\sum_{i\leq k}P_i,b])|+
\frac{1}{n}  |tr( \rho [\sum_{i=k+1}^n P_i,b])|.
\]
Firstly we can 
show
\beq\label{Teil}
|tr (\rho [\frac{1}{k}\sum_{i\leq k}P_i,b])|\leq r_{wk}
\eeq
 by
a slight modification of the arguments in the proof of Theorem \ref{GTheorem}:
Due to the fact, that for any qubit $i$ with $i\in \{1,\dots ,k\}$
 there is a step in the algorithm which
is the {\it last} access  on $i$ by 2-qubit-gates, every of those
  qubits is exposed to
the depolarizing channel, i.e., the state
$\rho$ can be written as $\rho =\tilde{D} (\nu)\otimes \mu$,
where $\nu$ and $\mu$ are states of the qubits $1,\dots,k$ and
$k+1,\dots,n$, respectively and $\tilde{D}$ is the $k$-fold depolarizing
channel on the qubits 
$1,\dots,k$. The map  $\tilde{D}$ is given by the restriction
of  $\prod_{i\leq k} D_i$ to this subsystem. 
By convexity arguments, it is sufficient to show
\[
|tr (((\tilde{D}\nu)\otimes \mu) [\frac{1}{k}\sum_{i\leq k} P_i,b])|\leq r_{wk}
\]
for every {\it pure} state $\nu$.
Since $\tilde{D}$ is the analogue to $D$ for the $k$-qubit-system
$1,\dots,k$,
the mean standard deviations of the  mixture
$\tilde{D}\nu$ can be estimated in analogy to 
 the  proof  of Theorem \ref{GTheorem}. 
Using Corollary \ref{gewichteteStandard} shows inequality (\ref{Teil}).

Since the qubits $k+1,\dots,n$ are still in a product state after the 
preparation 
procedure, the observables $\{P_i\}_{i\geq k+1}$ are stochastically 
independent. 
Therefore the variance of the sum $\sum_{i\geq k+1}P_i$ is the sum
of the variances of $P_i$.
Hence
the standard deviation of $\frac{1}{n}\sum_{i\geq k+1} P_i$
cannot exceed $\frac{1}{2n}\sqrt{n-k}$.
Hence we have
\[
|tr ( \rho [\frac{1}{n}\sum_{i\geq k+1}P_i,b])|\leq \frac{1}{n} \sqrt{n-k}.
\]
 The extension of the proof from a
pure  to a separable initial state is given by standard 
convexity arguments.  
\end{Proof} 

\begin{Corollary}
Modify the assumptions of Theorem \ref{Haupt} as follows:
Instead of 2-qubit-gates with depolarizing error we assume
a dephasing error, that is, we have imperfect 2-qubit-gates $g$ of
the form 
\[
g(\rho)=(1-w)u\rho u^*+w(M_i\circ M_j)(\rho),
\]
where $u$ is an arbitrary unitary acting on the qubits $i$ and $j$.

Then it is not possible to prepare a state outside
the slice defined by the hypersurfaces
\[
H_{\overline{P},b,\pm x}
\]
with $b$ and $x$ as in Theorem \ref{Haupt}.
\end{Corollary}

\begin{Proof}
In analogy to the proof of Theorem \ref{Haupt}:
Those qubits $1,\dots,k$, which are accessed at all, 
are subjected to the dephasing error, i.e., to the instrument
$G$  on the $k$-qubit Quantum Computer $1,\dots,k$.
\end{Proof}

Theorem \ref{main} shows up to which values the error rate $w$ of
a depolarizing channel has to be reduced in order to maintain 
macroscopic superpositions in the  sense discussed here:
In order to investigate the asymptotic behaviour of 
\[
r_{wn}=\sum_{l=0}^n
\frac{1}{\sqrt{l}}\sqrt{\frac{n-l}{n-1}} B_{nw}(l)+(1-w)^n
\]
we choose an arbitrary $0<\alpha <1$ and split the summation over $l$ into 2 
parts and get
\beqy
r_{wn}&=&\sum_{1\leq l \leq w\alpha n} \frac{1}{\sqrt{l}}\sqrt{\frac{n-l}{n-1}}
B_{nw}(l)
\\
&+&
\sum_{\alpha w  n < l \leq n}
\frac{1}{\sqrt{l}}\sqrt{\frac{n-l}{n-1}} B_{nw}(l) + (1-w)^n\\
&\leq&
\sum_{1\leq l\leq wn\alpha}  B_{nw}(l) +
\frac{1}{\sqrt{nw\alpha }} + (1-w)^n\\
&=&
\sum_{0\leq l \leq wn\alpha} B_{nw}(l)+ \frac{1}{\sqrt{nw\alpha}}
\eeqy
Due to the Tschebyscheff inequality the sum over the binomial coefficients
from $0$ to $wn\alpha$  is less
or equal to $\frac{1}{wn(1-\alpha)^2}$.
Hence we conclude for any $0<\alpha <1$:
\[
r_{wn}\leq \frac{1}{wn(1-\alpha)^2}+\frac{1}{\sqrt{nw\alpha}}.
\]

\section{Conclusions}

We have shown, in which way the parameter $e_\rho$
and the `hypersurface-criterion' 
can be used for detecting many-particle-entanglement.
States with large $e_\rho$ are fragile in the following sense:
We see that $r_{wn}$ is decreasing with $O(\sqrt{wn})$ for 
$n$ going  to infinity.
For a fixed error probability $w$ we conclude that
the maximum of $e_\rho$ which can be attained by states
subjected to the $n$-fold depolarizing channel is decreasing
with $O(1/\sqrt{n})$.

Furthermore we see
that the preparation of states $\rho$ 
 with a given $e_\rho$  by imperfect gates requires an  error probability
 decreasing with $1/n$ or faster.
Similarly, it requires an error probability not greater than $O(1/n)$
in order to have a state with a fixed  $e_\rho$ after one time step
if the decohering effect of the environment during one time step
is described by the map $D$.
Note that these are statements about the {\it physical} states
of the Quantum Computer in contrast to the {\it logical} states
which may be defined by a certain Quantum Code.

At the moment, we cannot see whether stronger bounds can be given.
However, it should be emphasized that our bounds for the fragility of
macroscopic superpositions are much weaker than those which
are suggested by simple but non-generic examples:
The instrument $D$ transduces the density matrix of the  cat state
$\frac{1}{\sqrt{2}}(|0\dots 0\rangle +|1\dots 1\rangle)$
to a density matrix $\rho$  containing the cat state with probability $(1-w)^n$
and containing unentangled states with probability $1-(1-w)^n$.
Therefore  we get exponential decay of $e_\rho$  for increasing $n$.
\end{multicols}

\section*{Acknoledgements}

Thanks to M. Grassl, M. R\"otteler and R. Schack for many correcting remarks.
This work was partially supported by grants of the project
`Quanten-Informationstheorie Karlsruhe-Innsbruck (QIKI)' of the State
of Baden-W\"urttemberg.


\begin{references}

\bibitem{SZL}
C.P. Sun, H. Zhan, and  X.F. Liu, technical report AMO 98-2, ITP-AC, 
 e-print quant-ph/9802029.

\bibitem{Un} W.G. Unruh, Phys. Rev. A {\bf 51} 993, (1995).
\bibitem{Di} D. Divincenzo and B. Terhal, Physics World, {\bf 53},
march 1998.
\bibitem{GJKKSZ}
D. Giulini, E. Joos, C. Kiefer, J. Kupsch, I.-O. Stamatescu, 
and H. D. Zeh, Decoherence and the Appearance of a Classical World
in Quantum Theory, Springer Verlag Berlin Heidelberg 1996. 
\bibitem{Zu1}
W.H. Zurek, Phys. Today {\bf 44}, 36, (1991).

\bibitem{Zu3} W.H. Zurek, Phys. Rev.  D {\bf 24}, 1516, (1981).

\bibitem{Pr} H. Primas, Chemistry, Quantum Mechanics and 
Reductionism, Springer Verlag Berlin 1981.
\bibitem{Zu2} W.H. Zurek, Progr. Theor. Phys. {\bf 89}, 281, (1993).

\bibitem{BBH} D. Braun, P. Braun, and F. Haake, e-print  quant-ph/9903040.

\bibitem{AB}
D. Aharonov and M. Ben-Or, 37th Annual Symposium on Foundations of 
Computer Science (FOCS) 46, (1996).



\bibitem{VPRK} V. Vedral, M.~B. Plenio, M.~A. Rippin and P.~L. Knight,
Phys. Rev. Lett., 1997  {\bf 98}, 2275 (1997).
\bibitem{SM1} J. Schlienz and G. Mahler, Phys. Lett. A {\bf 224}, 39, (1996). 
\bibitem{SM2} J. Schlienz and G. Mahler, Phys. Rev. A {\bf 52}, 4396, (1995).

\bibitem{AKN}
D. Aharonov, Alexei Kitaev, and Noam Nisan, STOC 1998.
\bibitem{JK} N. L. Johnson and S. Kotz, Urn models and their application, John
Wiley \& Sons, New York 1977.

\bibitem{St}
A. Steane, Phys. Rev. Lett. {\bf 77},  793, 1996 

\end{references}
\end{document}